\documentclass{article}

\usepackage{arxiv}

\usepackage[utf8]{inputenc} 
\usepackage[T1]{fontenc}    
\usepackage{hyperref}       
\usepackage{url}            
\usepackage{booktabs}       
\usepackage{amsmath}
\usepackage{amsfonts}       
\usepackage{nicefrac}       
\usepackage{microtype}      
\usepackage{cleveref}       
\usepackage{lipsum}         
\usepackage{graphicx}
\usepackage[numbers]{natbib}
\usepackage{doi}
\usepackage{makecell}

\usepackage{tikz}
\usepackage{pgfplots}

\title{Diff-Based Code Corruption using LLMs for Large-Scale Bugfix Benchmarking}

\date{}

\author{
        \href{https://orcid.org/0000-0002-0458-2576}{\includegraphics[scale=0.06]{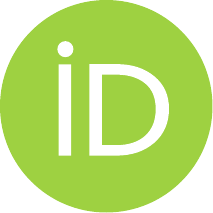}\hspace{1mm}Balázs Szalontai}\\
	\texttt{szalontaib@inf.elte.hu} \\
	\And
        \href{https://orcid.org/0009-0008-9782-3869}{\includegraphics[scale=0.06]{orcid.pdf}\hspace{1mm}Ábel Szauter}\\
	\texttt{szauterabel@inf.elte.hu} \\
        \And
        \href{https://orcid.org/0009-0000-1936-0207}{\includegraphics[scale=0.06]{orcid.pdf}\hspace{1mm}Balázs Márton}\\
	\texttt{g68pwv@inf.elte.hu} \\
        \And
        \href{https://orcid.org/0009-0002-5239-0517}{\includegraphics[scale=0.06]{orcid.pdf}\hspace{1mm}Péter Verebics}\\
	\texttt{e5cxij@inf.elte.hu} \\
        \And
        \href{https://orcid.org/0000-0003-3431-0667}{\includegraphics[scale=0.06]{orcid.pdf}\hspace{1mm}Balázs Pintér}\\
	\texttt{pinter@inf.elte.hu} \\
        \And
        \href{https://orcid.org/0000-0002-9503-9623}{\includegraphics[scale=0.06]{orcid.pdf}\hspace{1mm}Tibor Gregorics}\\
	\texttt{gt@inf.elte.hu}
}

\renewcommand{\headeright}{}
\renewcommand{\undertitle}{}
\renewcommand{\shorttitle}{Diff-Based Code Corruption using LLMs for Large-Scale Bugfix Benchmarking}

\hypersetup{
pdftitle={Diff-Based Code Corruption using LLMs for Large-Scale Bugfix Benchmarking},
pdfsubject={Large Language Model, Bugfix, Benchmark, Diff},
pdfauthor={Balázs Szalontai, Ábel Szauter, Balázs Márton, Péter Verebics, Balázs Pintér, Tibor Gregorics},
pdfkeywords={Large Language Model, Bugfix, Benchmark, Diff},
}

\begin{document}
\maketitle

\vspace{-1cm}
\begin{center}
	{
	\large
	Eötvös Loránd University\\Faculty of Informatics
	}
\end{center}
\vspace{2cm}

\begin{abstract}
  There are various benchmarks to evaluate bugfixing capabilities of Large
  Language Models. However, most widespread benchmarks do not fully reflect
  real-world bugfixing practices. They are small, weakening statistical
  reliability, and the buggy programs are often similar to one another,
  potentially distorting evaluation results. The range of bug types can also be
  narrow, failing to capture a representative range of bugs. To address these
  issues, we introduce MegaBugFix, a large-scale bugfixing benchmark containing
  12,629 buggy Python programs synthesized from correct ones by a Large Language
  Model. Bug injections were generated as diffs representing code changes.
  Through this approach, we were able to avoid common pitfalls of LLM-based
  mutation techniques like injecting overly simplistic bugs or failing to modify
  the input program. We evaluated 13 open-weight models on MegaBugFix and
  baseline benchmarks, finding consistently lower performance on MegaBugFix.
  This reveals that our benchmark presents more challenging bugs and exposes
  model failures that may remain hidden when evaluating on existing benchmarks.
  The benchmark and fine-tuned model used for bug injection are available at
  \href{https://hf.co/collections/szalontaib/megabugfix}{\path{hf.co/collections/szalontaib/megabugfix}}.
\end{abstract}


\newpage

\section{Introduction}

In recent years, deep learning-based methods have emerged to solve various
bugfixing tasks~\citep{yasunaga2020graphbased, coconut, sequencer, tbar}. Large
Language Models (LLMs) have also shown impressive bugfixing capabilities by
following instructions to identify and correct code
errors~\citep{granite2024granite, jiang2024codecomparisontuningcode,
jiang2024cursorcoreassistprogrammingaligning,
muennighoff2024octopackinstructiontuningcode}. Multiple benchmarks have been
introduced to measure and compare the bugfixing performance of these methods,
with QuixBugs~\citep{lin2017quixbugs} and
HumanEvalFix~\citep{muennighoff2024octopackinstructiontuningcode} being two
notable examples. Such benchmarks typically consist of a set of incorrect
programs that the model must repair. The repaired programs are then evaluated
using test cases. The benchmark scores are usually calculated using the pass@k
metric~\citep{chen2021evaluatinglargelanguagemodels}, which measures the
proportion of problems for which at least one out of k generated fixes is
successful. Most commonly, the pass@1 score is used, which represents the
proportion of successfully fixed programs in a single attempt.

Although the most utilized benchmarks are widely adopted and are useful for
standardized evaluation, they do not necessarily represent actual bugfixing
performance in practical settings. Such benchmarks might rely on a limited set
of buggy programs (e.g. 164 samples in HumanEvalFix and 40 in QuixBugs).
Furthermore, they often consist of collections of similar samples in terms of
structure and complexity. For instance, in both HumanEvalFix and QuixBugs, each
program corresponds to a single, relatively short function. Moreover, the nature
of bugs in many existing datasets tends to be narrow in scope and localized to a
specific line or code section. For example, in QuixBugs, each bug is limited to
a single line, whereas in HumanEvalFix, bugs involve an average of 1.1 changed
lines. This limited diversity restricts their ability to capture the broad
spectrum of issues that arise in real-world software projects.

The primary bottleneck in creating a bugfixing benchmark is dataset creation. It
is a significant effort to manually produce large numbers of buggy programs,
their corresponding correct versions, and test cases to evaluate generated
fixes. While many public datasets provide correct program implementations with
accompanying tests~\citep{lai2022ds1000naturalreliablebenchmark,
austin2021programsynthesislargelanguage, algorithms, petersen2022dataset}, they
generally lack buggy counterparts, making them unsuitable as bugfixing benchmark
datasets. 

One way to alleviate this problem, pioneered by
DebugBench~\citep{tian2024debugbench}, is to create synthetic datasets by
introducing bugs into bug-free programs using LLMs. Building on this idea,
CodeEditorBench~\citep{guo2025codeeditorbench} and DebugEval~\citep{yang2025coast}
also leveraged this approach. However, these methods are prone to generate
syntax-level bugs that are easier to detect and fix, and not deeper,
semantic-level bugs. For example, more than half of the buggy Python programs in
DebugBench (832/1414) are not parsable by the Python interpreter, and a similar
proportion of unparsable Python code appears in CodeEditorBench's Primary
(312/716) and Plus (193/356) datasets. A large-scale dataset with deeper bugs is
needed, where the program remains syntactically correct. We believe that
fine-tuning is a viable path to do this.


LLMs used in a code editing scenario can be prone to produce outputs that are
identical to the input~\citep{li2024instructcoderinstructiontuninglarge,
brownlee2025large, munson2024out}, which therefore also complicates LLM-based
code corruption. A potential solution is to generate the \emph{diff}
representing the code change rather than producing the full output
directly~\citep{muennighoff2024octopackinstructiontuningcode, bo2023cct5}. To the
best of our knowledge, this approach has not yet been explored for code mutation
or program corruption. We hypothesize that fine-tuning a model to generate diffs
instead of the final output provides a reliable mechanism to inject bugs while
ensuring the code is actually modified.

In this paper, we introduce \textit{MegaBugFix}, a large-scale benchmark to
measure bugfixing capabilities. By fine-tuning an open-weight LLM for the task
of program corruption via diff generation, we automatically inject bugs into
programs, resulting in a large-scale benchmark dataset of 12,629 incorrect
programs. To ensure dataset diversity, programs were gathered from six different
sources, and they were corrupted in multiple ways. Alongside the corrupted
programs, we also gathered validating test cases, which are used to evaluate
correctness of the fixed programs. To facilitate straightforward evaluation, we
provide a framework for the proposed benchmark that includes a unified test
execution suite for running test cases that originate from different sources.

\section{Related Work}\label{sec:related}

Bugfixing benchmarks have been in use before the rise of LLMs, including
Defects4J from 2014~\citep{just2014defects4j}, which contains 357 real bugs from
5 open source Java programs. Another notable benchmark from this era is
QuixBugs~\citep{lin2017quixbugs}, containing 40 buggy programs, and test cases to
validate the fixes. Although programs in the QuixBugs benchmark can be used to
evaluate bugfixing capabilities, they are very limited in number (with only 40
programs), and they represent a narrow range of problems with only one
single-line bug in each of them.

CodeXGLUE~\citep{lu2021codexgluemachinelearningbenchmark} is a set of benchmarks
which includes 10 tasks to evaluate and compare models. One of these tasks is
code repair, which utilizes the Bugs2Fix dataset~\citep{tufano2019empirical}.
The goal is to evaluate bugfixing performance on Java programs by comparing the
fixed programs to ground truths using exact match accuracy, BLEU, and CodeBLEU
metrics. While this benchmark dataset contains bugs from real-world projects,
the utilized metrics might not effectively reflect actual bugfixing
performance, as they rely on syntactical similarity rather than program
execution.

Another benchmark to evaluate Python bugfixing capabilities is
BugsInPy~\citep{bugsinpy}, which includes 493 real-world bugs from 17 Python
projects across diverse domains, such as machine learning, developer tools,
scientific computing, and web frameworks. To ensure high-quality bugs,
repositories are obtained from GitHub, each with more than 10,000 stars. An
improved version of this dataset~\citep{bugsinpy2} extends and refines the
original benchmark.

Since the introduction of the HumanEvalPack family of
benchmarks~\citep{muennighoff2024octopackinstructiontuningcode}, HumanEvalFix has
become widely used to evaluate bugfixing capabilities. It utilizes buggy
variants of the well-known code generation benchmark,
HumanEval~\citep{chen2021evaluatinglargelanguagemodels}. The evaluated model is
prompted to follow the instructions to fix bugs in these buggy variants. The
benchmark result indicates the number of successfully generated (repaired)
programs, similarly to the original HumanEval benchmark. This benchmark has
become one of the standards in the literature, which facilitates straightforward
comparison~\citep{granite2024granite,
jiang2024cursorcoreassistprogrammingaligning, chae2024coffee,
jiang2024codecomparisontuningcode, singhal2024nofunevalfunnycodelms}.

To systematically evaluate the debugging capabilities of LLMs, Tian et al.
introduced DebugBench~\citep{tian2024debugbench}, a large-scale benchmark
comprising 4,253 instances across C++, Java, and Python. The dataset covers
four major bug categories and 18 minor types, including missing colons,
condition error, faulty indexing, and unclosed string literals. The dataset was
created by injecting bugs into LeetCode code snippets using GPT-4, followed by
manual validation. This approach was also leveraged by Gou et al., who
introduced CodeEditorBench~\citep{guo2025codeeditorbench}, covering code
translation, polish, and requirement switching tasks alongside code debug.
Furthermore, the approach and data from DebugBench was also utilized by Yang et
al.~\citep{yang2025coast}, who proposed DebugEval, a benchmark for evaluating
the debugging capabilities of LLMs by emulating the multi-stage human debugging
process.

An alternative to LLM-based code corruption for benchmark construction could be
using rather traditional approaches. Ouyang et al. propose MuBench and use it to
investigate several automated program repair tools
\citep{ouyang2024benchmarking}. This benchmark uses Defects4J programs that have
been modified with the Major mutation testing framework \citep{rene2014major} in
multiple ways. Their benchmark consists of 100 mutated samples for 17 projects,
resulting in a benchmark dataset of 1,700 buggy programs.

To address the gap of diverse language support for automatic program repair
tools, Liu et al.\ introduced
MdEval~\citep{liu2024mdevalmassivelymultilingualcode}, a comprehensive
multilingual debugging benchmark covering 20 programming languages. The
benchmark contains 1,299 human-annotated buggy programs, evaluated across three
tasks (bug identification, localization, and program repair), resulting in a
total dataset size of almost 3,600. They also provide a leaderboard, which
facilitates comparison between models in multilingual debugging scenarios.
Furthermore, the authors released MdEval-Instruct, a separate dataset generated
via automatic bug injection, which was used to train models later evaluated on
MdEval.

Although not specifically focused on creating bugfix benchmarks, some recent
work examines the ability of LLMs to generate code mutations and synthetic bugs.
Khanfir et al.\ introduced $\mu$Bert~\citep{degiovanni2022mubert}, which relies
on CodeBERT~\citep{feng2020codebert} to generate realistic code mutations through
token-level replacements, enabling mutation testing without fine-tuning. Tip et
al.\ proposed LLMorpheus~\citep{tip2025llmorpheus}, a tool that leverages LLMs to
inject context-aware mutants into code, producing bugs that better resemble real
bugs than traditional operator-based methods. Wang et al.\ analyzed the code
mutation capabilities of LLMs~\citep{wang2025comprehensivestudylargelanguage}
across two Java benchmarks (Defects4J~\citep{just2014defects4j} and
ConDefects~\citep{wu2024condefects}). They show that although LLMs can create
diverse mutations that are behaviorally closer to real bugs, they also have
worse compilability rate, useless mutation rate, and equivalent mutation rate
than those generated by rule-based approaches. In our work, we observe a similar
caveat, which we detail in \autoref{sec:corrupting_programs}. Ibrahimzada et al.
presented BugFarm~\citep{ibrahimzada2025challengingbugpredictionrepair}, a
framework for generating complex synthetic bugs with LLMs, focusing on
hard-to-detect and hard-to-repair defects to challenge bugfixing capabilities of
Transformer-based models. Jasper et al. present the BugGen pipeline
\citep{jasper2025buggen} that uses a multi-agent LLM architecture to
automatically generate software bugs. It uses specialized agents to analyze code
context and inject complex defects that mimic human errors.

Some studies have investigated the ability of LLMs to handle code modifications
formulated as diffs rather than as separate input and output programs. Li et al.
presented CodeReviewer~\citep{li2022automating}, a pre-trained encoder-decoder
model designed for code review tasks. It addresses problems such as code change
quality estimation and automatic code review generation by representing code
changes as diffs, making it the first pre-trained model to leverage code diffs
as input in a code review setting. Their framework can also refine (improve) the
original programs using the review comments.

Lin et al.\ introduced CCT5~\citep{bo2023cct5}, which is also a pre-trained model
designed for code change tasks. Their pre-training tasks include a natural
language to programming language generation objective, in which CCT5 learns to
generate newly added code lines based on a masked code diff that represents the
original code, and a commit message that describes the intended change. Their
experiments demonstrate that CCT5 outperforms contemporary models (such as
CodeReviewer and CodeT5~\citep{wang2021codet5}) on three widely studied code
change tasks (commit message generation, just-in-time comment update,
just-in-time defect prediction) and two code review-related tasks (code review
generation, code change quality estimation).

Muennighoff et al.\ explored LLM-based bugfixing via diff
generation~\citep{muennighoff2024octopackinstructiontuningcode} by fine-tuning a
model to follow a line diff format for fixes. They fine-tuned
SantaCoder~\citep{allal2023santacoderdontreachstars} on a subset of CommitPackFT,
which is a dataset containing programs before and after commits along with
corresponding commit messages. This approach yielded better bugfixing
performance compared to the original SantaCoder model, as well as to SantaCoder
fine-tuned on commits using a standard full code generation setting. Fan et al.
conducted an empirical study~\citep{fan2025exploring} on the use of LLMs for code
change-related tasks, namely code review generation, commit message generation,
and just-in-time comment updates. They also investigated whether LLMs perform
better when the LLM input is provided as a diff rather than as two separate code
snippets. Their results showed that diffs make it easier for LLMs to identify
changes, leading to better performance.

\section{The Proposed Benchmark}

Our proposed benchmark consists of 12,629 buggy programs, their correct
counterparts, and a framework that can be used to evaluate the bugfixing
performance of LLMs. We used 6 publicly available datasets to collect correctly
implemented programs, which we corrupted in multiple ways. The corruption was
carried out by an LLM specifically fine-tuned for the task of bug injection.
Alongside the correct programs, validating test cases were also gathered and
unified to evaluate correctness of the fixed programs.

Here, we first outline how we obtain the training dataset, used for fine-tuning.
Second, we describe the approach and parameters of fine-tuning the LLM. Third,
we summarize the source of correct programs and their validating test cases, and
also outline our approach of creating a unified evaluation framework. Then we
present the method of applying our fine-tuned LLM on the correct programs to
obtain their incorrect variants.

\subsection{Fine-Tuning Dataset}

In order to obtain a large number of buggy programs from correctly implemented
ones, we fine-tune an LLM to introduce bugs in ways that mirror typical mistakes
made by humans. The first step of this process is to create the training
dataset. We use human-written programs, containing both correct and corrupted
versions of the same code. We source these from Project
CodeNet~\citep{puri2021codenetlargescaleaicode}, which is a large-scale dataset
containing almost 14 million submissions to a total of 4,053 coding problems.
The submission verdicts are also known, through which we can obtain
\texttt{(correct, buggy)} pairs of Python programs. These pairs are obtained by
first grouping by tasks and users, and then separating the submissions that are
(i) accepted and (ii) rejected due to wrong answer. If the two programs in a
pair of \texttt{(accepted, rejected)} submission are similar (with at least 70\%
similarity\footnote{Similarity is measured using the rapidfuzz
ratio~\citep{maxbachmann2024}.}), they are included in the training dataset as a
\texttt{(correct, buggy)} pair. 

Initial experiments with non-finetuned LLMs have shown that they are generally
not well-suited for synthesizing good-quality corrupted alternatives directly
from a correct program. In several cases, models introduced an overly simplistic
and easily noticeable bug. Although our initial experiments mostly focused on
small open models, this behavior can be also observed in case of GPT-4: some of
the bugs in DebugBench~\citep{tian2024debugbench} are unrealistic, such as
variable names and values that explicitly reference the bug itself (e.g.,
\texttt{unclosedString = "bug introduction}), malformed expressions that
literally contain the word bug (e.g. \texttt{return
Math.max(a.length(),b.<bug>null}), or comments that directly indicate the bug
(e.g., \texttt{// Here is the bug}). To avoid such issues, we initially
attempted to fine-tune LLMs for the task of bug injection in an input-output
manner, but the models were still mostly underperforming, generating unmodified
programs in many cases.

We overcome these shortcomings by fine-tuning an LLM to synthesize the
\textit{diff} between the correct and buggy programs instead of the buggy
program itself. Such a diff includes the inserted and deleted lines (marked with
a leading ``+'' and ``-''), as well as the unchanged lines. We found that LLMs
perform better in bug injection using this format, compared to having them
generate the full buggy program.

In order to ensure the quality of the training dataset, we first formatted each
program with the \emph{black} formatter~\citep{LangaBlackTheuncompromising}.
This is done to prevent unnecessary changes included in the diff caused by
formatting only. We then filter the code pairs, resulting in a final dataset of
10,310 pairs that fulfill the following criteria:
\begin{itemize}
    \item The diff includes both insertions and deletions
    \item The diff is not too large. A diff is considered acceptable if
    it meets the following condition based on its length: $L_{\text{diff}} \le
    \tfrac{1}{4}(L_{\text{orig}} + L_{\text{mod}})$, \\
    where $L$ represents the length of deleted and inserted lines
    ($L_{\text{diff}}$), length of the original file ($L_{\text{orig}}$), and
    length of the modified file ($L_{\text{mod}}$).
    \item The diff does not contain the substring ``import''
    \item The inserted content and deleted content are similar enough ($>
    60\%$) but not too similar ($< 80\%$). Before measuring similarity, the
    characters of both inserted and deleted lines get sorted, in order to avoid
    modifications that just rearrange commutative parts of the code without
    modifying the behavior (such as $a+b \to b+a$). 
\end{itemize}

\subsection{Fine-Tuning to Inject Bugs}

WizardCoder-13B-Python~\citep{luo2023wizardcoderempoweringcodelarge} is an
open-weight model that excels at following instructions related to Python,
making it well-suited for code transformation objectives. The goal is to turn
this model into one that maps correct code to a diff that represents its
corruption. The training data is formatted into pairs of correct programs and
diffs, which we parse into the following format: 
\texttt{[PYTHON] \dots\ [/PYTHON] [DIFF] \dots\ [/DIFF]}.

We used LoRA~\citep{hu2021loralowrankadaptationlarge} with every
layer of the network selected as a target for fine-tuning. The dimension of
the low-rank matrices was set to \boldmath$r = 512$, and the scaling factor for
the weight matrices was set to \boldmath$\alpha = 1024$. The learning rate was
set to \boldmath$\eta = 2 \cdot 10^{-4}$. To preserve more information, we
chose a low dropout probability of \boldmath$p = 0.1$.

We fine-tuned WizardCoder for multiple epochs using early stopping, with a
modified version of the QuixBugs benchmark serving as the validation dataset.
We took the reference solutions (the bug-free programs) of the benchmark
dataset, and fed it to the network after each epoch. Since the goal of our
model is to corrupt as many programs as possible, we stopped training once
there was any increase in the number of programs that pass all its test cases.

\subsection{Gathering Correct Programs with Validating Test Cases}\label{sec:gathering}

To create the benchmark dataset, we rely on existing, correctly implemented
programs accompanied by validating test cases. Several benchmarks and datasets
are suitable for our purposes, including code generation benchmarks and
submissions to automatically graded assignments. We select 6 sources to gather
correct program implementations: we use the canonical solutions of four
benchmarks (namely HumanEval~\citep{chen2021evaluatinglargelanguagemodels},
QuixBugs~\citep{lin2017quixbugs},
DS-1000~\citep{lai2022ds1000naturalreliablebenchmark} and
MBPP~\citep{austin2021programsynthesislargelanguage}), a dataset of student
solutions to programming assignments about algorithms and data
structures~\citep{petersen2022dataset}, and the Python subset of the GitHub
repository named The Algorithms~\citep{algorithms}. We selected these sources to
capture a wide spectrum of programming styles, domains, and difficulty levels.
The collected programs include concise, function-level tasks (QuixBugs,
HumanEval, MBPP), diverse algorithmic problem solutions (from The Algorithms
repository), student-written submissions to academic assignments (from the
student submissions dataset), and data science–related programs (DS-1000).

Validating the correctness of programs with test cases is performed differently
across the different datasets. To address this, we introduce a unified framework
for executing test cases. Our framework relies on pytest~\citep{pytest} to
collect and run tests. As the original sources used different methods for test
execution, they had to be adapted to our unified framework. In The Algorithms
GitHub repository, tests were written as doctest cases, so we collected them and
converted them into pytest classes. For the dataset of student solutions to
assignments, test cases were generated from a CSV file containing the test
specifications. In the DS-1000 benchmark, tests were stored in a JSONL file,
which we parsed and transformed into pytest test cases. The tests for the
HumanEval and MBPP benchmarks were generated in a similar way from the benchmark
data. For QuixBugs, we used the original test code and transformed it into
parametrized pytest test functions. We further modified QuixBugs tests for added
safety by deep copying test data to eliminate side effects when testing multiple
solutions.

LLM generated code can be unreliable. To ensure a safe and consistent execution
environment, we created a Docker image, allowing users to run the tests inside a
container. We also introduce some additional measures to ensure test runs are
consistent and deterministic, such as seeding random generators with a fixed
seed. Additionally, we introduce a timeout limit to solution modules to avoid
hanging pytest indefinitely during the collection phase. Furthermore, we limit
the memory used by running each test to 2GB: if a test exceeds this limit, the
process is killed, and the test is marked as failed.

\subsection{Corrupting Programs}\label{sec:corrupting_programs}

Once the model is fine-tuned, the corrupted programs need to be synthesized by
injecting bugs into the correct programs. In order to do this, we feed the
correct programs to the fine-tuned model in the following format:
\texttt{[PYTHON] \dots\ [/PYTHON] [DIFF]}. Then, we have the model generate until
the \texttt{[/DIFF]} separator. The generated diff is extracted as the result,
which is then applied to obtain the corrupted program.

Since it is possible to corrupt a program in multiple ways (by injecting
different bugs), we use sampling with a temperature of 0.5 to obtain multiple
corrupted versions of the same program. For each program, 10 samples are
generated and then passed through a filtering process. We remove duplicate
generations, unparsable generations, those that are identical to the input,
generations that are not buggy (i.e., passed all test cases), and those that are
overly modified (string similarity of less than 80\% compared to the original).

Through this process, we obtain a variable number of corrupted variants for each
program. These generations form the core of MegaBugFix and serve as inputs to
the evaluated model.\ \autoref{tab:datasets} summarizes the sources of programs
(originating from the 6 datasets described in \autoref{sec:gathering}), the size
of each dataset, and the number of corrupted programs from each dataset.

\begin{table}[t!]
    \centering
    \caption{Summary of the datasets comprising MegaBugFix}
    \begin{tabular}{l|c|c}
        \textbf{Dataset origin} & \textbf{Original size} & \textbf{Corrupted size} \\ \hline\hline
        HumanEval~\citep{chen2021evaluatinglargelanguagemodels}  & 164  & 648 \\\hline
        QuixBugs~\citep{lin2017quixbugs}                         & 40   & 153 \\\hline
        DS-1000~\citep{lai2022ds1000naturalreliablebenchmark}    & 1000 & 3537 \\\hline
        MBPP~\citep{austin2021programsynthesislargelanguage}     & 974  & 3336 \\\hline
        The Algorithms~\citep{algorithms}                        & 607  & 2003 \\\hline
        AD submissions~\citep{petersen2022dataset}               & 653  & 2952 \\\hline\hline
        \textbf{Total}                                          & 3438 & 12629
    \end{tabular}
    \label{tab:datasets}
\end{table}

As shown in \autoref{tab:datasets}, the corrupted programs are generated from
3,438 correct programs. Although we initially generate 10 corrupted programs for
each correct program, there is only a 3.67x increase in the dataset size. This
discrepancy can be explained by the filtering process, which removes a large
fraction of sampled corruptions, as they are unparsable, duplicated, identical
to the original, not actually buggy, or too dissimilar. This observation aligns
with the findings of Wang et al., who also report high rates of non-compilable,
duplicate, and equivalent mutants among LLM-generated
corruptions~\citep{wang2025comprehensivestudylargelanguage}.

\begin{figure}[t!]
    \centering
    \includegraphics[width=\textwidth]{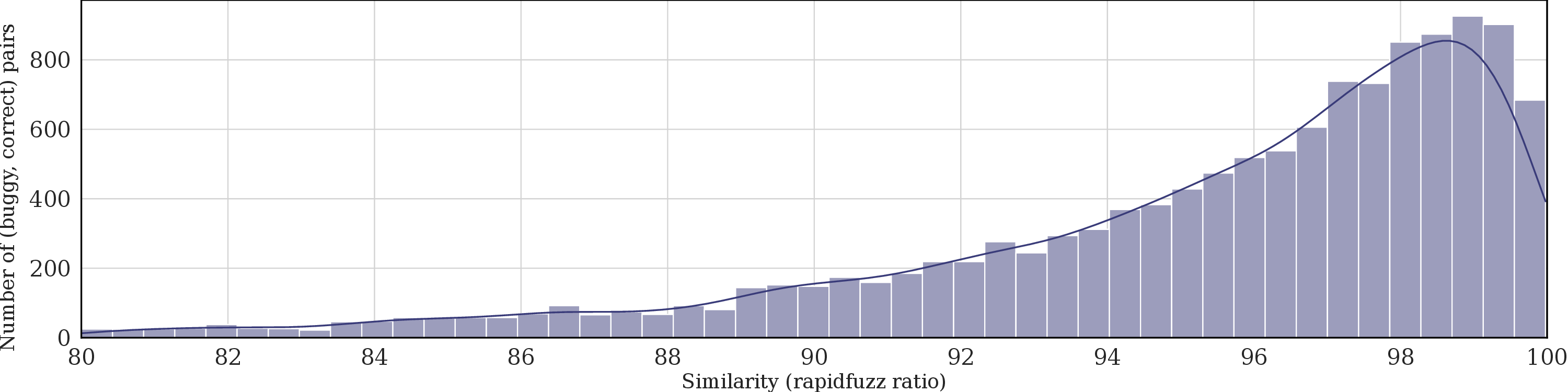}
    \caption{Histogram of string similarity ratios between correct and
    corrupted programs in the benchmark dataset. The large number of pairs with
    similarity close to 100\% indicates that a large portion of the corruptions
    are local and include only a subtle change in the code.}
    \label{fig:similarity-histogram}
\end{figure}

\autoref{tab:stats} provides statistics on the benchmark dataset to highlight
how the corrupted programs differ from their fixed versions. We report average,
median, minimum and maximum values for the following metrics: number of lines in
correct and corrupted programs, number of functions in corrupted programs,
number of differing lines (including data on insertions and deletions
separately), and similarity between correct and corrupted programs. We
illustrate the distribution of similarities between correct and corrupted
programs on a histogram in \autoref{fig:similarity-histogram}.

\begin{figure}[b!]
    \centering
    \includegraphics[width=\textwidth]{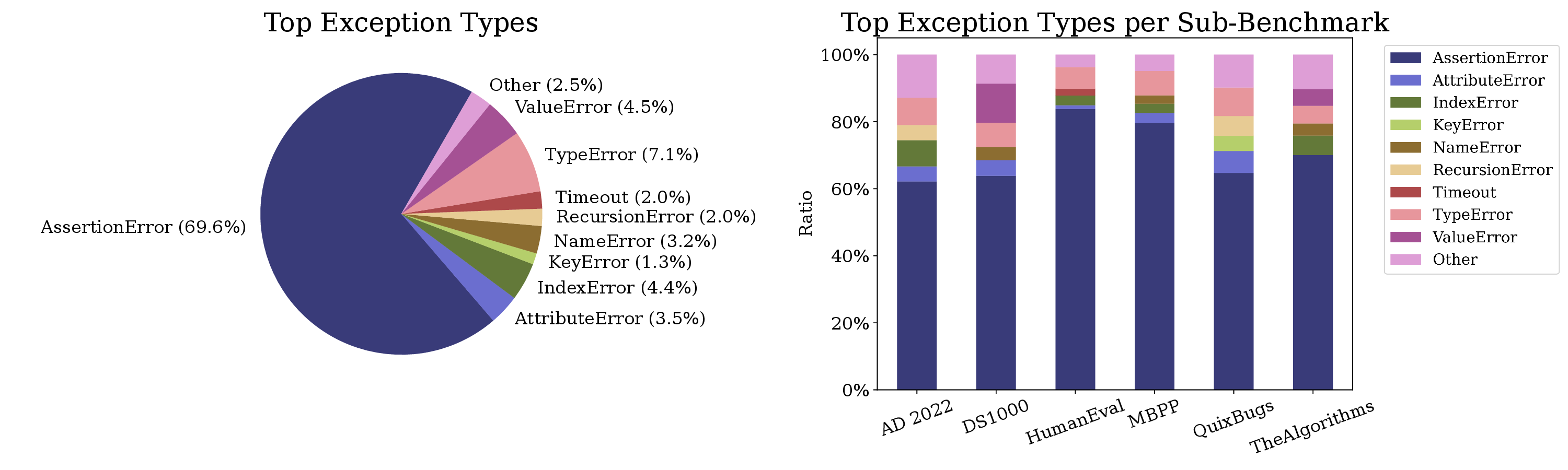}
    \caption{Exception types raised by the corrupted programs in the benchmark
    dataset. As most of the injected bugs represent semantic errors, the most
    common exception type is AssertionError. This is followed by other types of
    errors, such as TypeError or ValueError.}
    \label{fig:exception-types}
\end{figure}

Furthermore, we characterize the types of bugs present in the benchmark dataset.
To do this, we extract the exception types raised by the corrupted programs when
executing their test cases. The most common exception type is AssertionError,
which is raised when the output of the corrupted program does not match the
expected output. This is followed by other types of errors, such as
TypeError or ValueError.\ \autoref{fig:exception-types} shows the
distribution of these exception types.

\begin{table}[b!]
    \centering
    \caption{Statistics of the MegaBugFix benchmark}
    \begin{tabular}{l|c|c|c|c}
        \textbf{Metric} & \textbf{Mean} & \textbf{Median} & \textbf{Min} & \textbf{Max} \\ \hline\hline
        Correct program length & 12.97 & 11 & 2 & 156 \\ \hline
        Corrupted program length & 12.84 & 10 & 2 & 148 \\ \hline
        Number of functions & 1.47 & 1 & 1 & 17 \\ \hline
        Modified lines & 4.28 & 3 & 1 & 50 \\ \hline
        Added lines & 2.08 & 1 & 0 & 38 \\ \hline
        Removed lines & 2.21 & 2 & 0 & 39 \\ \hline
        String similarity ratio (\%) & 95.32\% & 96.58\% & 80.0\% & 99.98\%
    \end{tabular}
    \label{tab:stats}
\end{table}

\section{Evaluation}

In this section, we perform two experiments. First, we evaluate how current
open-weight models perform on the MegaBugFix benchmark. We contextualize these
findings by providing results from other established benchmarks to understand
how performance on MegaBugFix relates to them. In the second experiment, we
fine-tune LLMs for bugfixing using our benchmark dataset. We believe that if the
buggy programs in our benchmark are genuinely representative of real-world bugs,
this fine-tuning should improve the bugfixing ability of language models.

\subsection{Evaluating LLMs on the Benchmark}\label{sec:comparison}

\begin{figure*}[b!]
    \centering

    \begin{tikzpicture}
        \begin{axis}[
            ybar,
            bar width=3pt,
            ymin=0, ymax=105,
            ytick={0,10,20,30,40,50,60,70,80,90,100},
            height=6cm,
            width=1\linewidth,
            enlarge y limits=0,
            enlarge x limits=0.08,
            symbolic x coords={
                Qwen2.5-Coder-0.5B,
                Qwen2.5-Coder-1.5B,
                Qwen2.5-Coder-3B,
                Qwen2.5-Coder-7B,
                Qwen2.5-Coder-14B,
                Qwen2.5-Coder-32B,
                Yi-Coder-1.5B,
                Yi-Coder-9B,
                DeepSeek-Coder-1.3b,
                DeepSeek-Coder-6.7b,
                CodeLlama-7b,
                CodeLlama-13b,
                CodeGemma-7b
            },
            xtick=data,
            xticklabel style={
                rotate=20,
                anchor=east,
                font=\tiny,
                xshift=1.3em,
                yshift=-0.5em
            },
            ylabel={Score (\%)},
            ymajorgrids=true,
            y grid style={dashed, gray!30},
            axis background/.style={fill=gray!10, opacity=0.5},
            nodes near coords,
            every node near coord/.append style={font=\tiny, yshift=1pt},
            point meta=explicit symbolic,
            legend style={
                at={(0.48,1.18)},
                anchor=north,
                legend columns=-1,
                font=\normalsize
            },
            area legend,
            every node near coord/.append style={font=\tiny, yshift=0.5pt},
        ]

        \addplot[
            style={purple, fill=purple, opacity=0.6},
            every node near coord/.style={
                font=\tiny,
                rotate=90,
                xshift=11pt,
                yshift=0.22em,
            },
        ]
        coordinates {
            (Qwen2.5-Coder-0.5B,37.70) [37.70\%]
            (Qwen2.5-Coder-1.5B,49.18) [49.18\%]
            (Qwen2.5-Coder-3B,52.46) [52.46\%]
            (Qwen2.5-Coder-7B,67.21) [67.21\%]
            (Qwen2.5-Coder-14B,80.33) [80.33\%]
            (Qwen2.5-Coder-32B,85.25) [85.25\%]
            (Yi-Coder-1.5B,44.26) [44.26\%]
            (Yi-Coder-9B,52.46) [52.46\%]
            (DeepSeek-Coder-1.3b,37.70) [37.70\%]
            (DeepSeek-Coder-6.7b,73.77) [73.77\%]
            (CodeLlama-7b,39.34) [39.34\%]
            (CodeLlama-13b,52.46) [52.46\%]
            (CodeGemma-7b,60.66) [60.66\%]
        };

        \addplot[
            style={orange, fill=orange, opacity=0.6},
            every node near coord/.style={
                font=\tiny,
                rotate=90,
                xshift=10pt,
                yshift=-0.27em,
            },
        ]
        coordinates {
            (Qwen2.5-Coder-0.5B,25.0) [25.0\%]
            (Qwen2.5-Coder-1.5B,35.0) [35.0\%]
            (Qwen2.5-Coder-3B,55.0) [55.0\%]
            (Qwen2.5-Coder-7B,62.5) [62.5\%]
            (Qwen2.5-Coder-14B,65.0) [65.0\%]
            (Qwen2.5-Coder-32B,75.0) [75.0\%]
            (Yi-Coder-1.5B,65.0) [65.0\%]
            (Yi-Coder-9B,65.0) [65.0\%]
            (DeepSeek-Coder-1.3b,45.0) [45.0\%]
            (DeepSeek-Coder-6.7b,65.0) [65.0\%]
            (CodeLlama-7b,47.5) [47.5\%]
            (CodeLlama-13b,27.5) [27.5\%]
            (CodeGemma-7b,55.0) [55.0\%]
        };

        \addplot[
            style={blue, fill=blue, opacity=0.6},
            every node near coord/.style={
                font=\tiny,
                rotate=90,
                xshift=11pt,
                yshift=-0.8em,
            },
        ]
        coordinates {
            (Qwen2.5-Coder-0.5B,12.20) [12.20\%]
            (Qwen2.5-Coder-1.5B,35.37) [35.37\%]
            (Qwen2.5-Coder-3B,40.85) [40.85\%]
            (Qwen2.5-Coder-7B,58.54) [58.54\%]
            (Qwen2.5-Coder-14B,41.46) [41.46\%]
            (Qwen2.5-Coder-32B,70.73) [70.73\%]
            (Yi-Coder-1.5B,26.83) [26.83\%]
            (Yi-Coder-9B,60.37) [60.37\%]
            (DeepSeek-Coder-1.3b,20.12) [20.12\%]
            (DeepSeek-Coder-6.7b,49.39) [49.39\%]
            (CodeLlama-7b,20.12) [20.12\%]
            (CodeLlama-13b,18.90) [18.90\%]
            (CodeGemma-7b,43.90) [43.90\%]
        };

        \addplot[
            style={green!60!black, fill=green!60!black, opacity=0.6},
            every node near coord/.style={
                font=\tiny,
                rotate=90,
                xshift=11pt,
                yshift=-1.3em,
            },
        ]
        coordinates {
            (Qwen2.5-Coder-0.5B,11.31) [11.31\%]
            (Qwen2.5-Coder-1.5B,11.96) [11.96\%]
            (Qwen2.5-Coder-3B,20.87) [20.87\%]
            (Qwen2.5-Coder-7B,35.12) [35.12\%]
            (Qwen2.5-Coder-14B,43.14) [43.14\%]
            (Qwen2.5-Coder-32B,47.47) [47.47\%]
            (Yi-Coder-1.5B,20.44) [20.44\%]
            (Yi-Coder-9B,35.80) [35.80\%]
            (DeepSeek-Coder-1.3b,18.43) [18.43\%]
            (DeepSeek-Coder-6.7b,26.91) [26.91\%]
            (CodeLlama-7b,9.34) [9.34\%]
            (CodeLlama-13b,11.85) [11.85\%]
            (CodeGemma-7b,16.50) [16.50\%]
            
        };

        \legend{
            MdEval (Python bugfix),
            QuixBugs,
            HumanEvalFix,
            MegaBugFix
        }

        \end{axis}
    \end{tikzpicture}
    \caption{Performance of open-weight LLMs on MegaBugFix. MdEval (Python
    bugfixing subset), QuixBugs and HumanEvalFix results are also included for
    comparison. MegaBugFix results serve as baselines for our proposed
    benchmark. All evaluations were conducted locally. Although we used
    instruction-tuned models for our evaluations, the ``Instruct'' or ``Chat''
    substring from model names is omitted.}
    \label{fig:results}
\end{figure*}
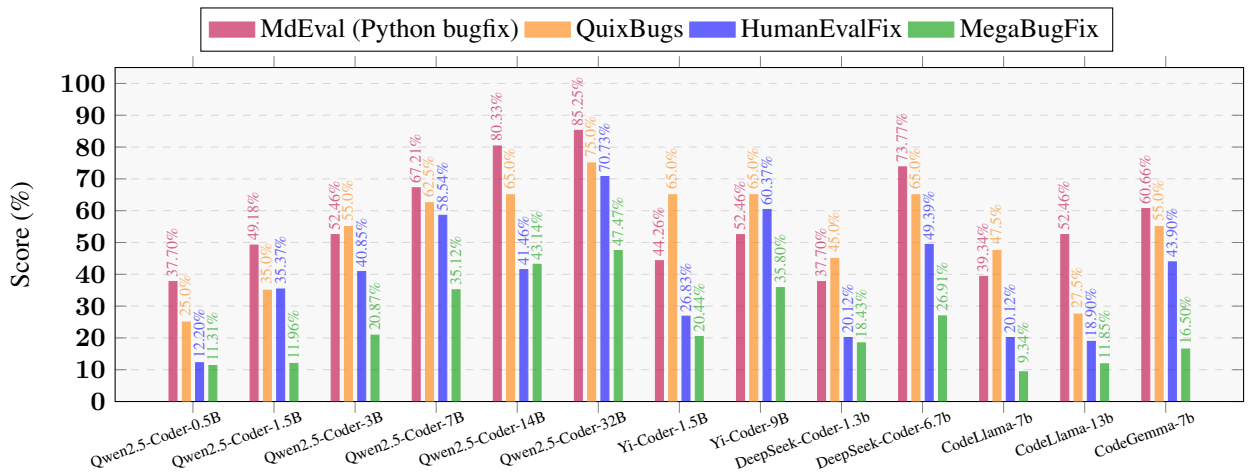

We provide MegaBugFix benchmark results for well-known open-weight LLMs. These
results could serve as baselines for the future, facilitating comparison with
newly published models. Here, we focus on smaller open-weight models that were
trained to excel at software engineering tasks. Alongside MegaBugFix results, we
provide HumanEvalFix, QuixBugs and MdEval (its Python bugfixing subset) results
for comparison. The evaluated models and their corresponding benchmark results
can be seen in \autoref{fig:results}.

All evaluation results provided in this paper originate from local evaluations.
To obtain HumanEvalFix and QuixBugs scores, the Bigcode LM Evaluation Harness
framework~\citep{bigcodeevaluationharness} was utilized. To evaluate models on
MdEval, we prompted the models with the provided instructions and used the test
cases to validate correctness of the generated programs. In evaluations, we
measured the pass@1 performance using greedy decoding. The models were loaded
with bf16 precision. The prompt template format was set to the format suggested
by the respective model authors. We set the generation length thresholds to be
sufficiently high so that a reasonably sized output fits comfortably. The
maximum length of generation in case of HumanEvalFix and QuixBugs was set to
2048. For MegaBugFix, the maximum number of generated tokens was set to 4096,
while for MdEval it was set to 1024.

\subsection{Improving LLMs by Fine-Tuning on the Benchmark Dataset}

Although the LLM used for bug injection was fine-tuned to mimic real program
corruptions, it is not immediately clear whether the resulting corrupted
programs contain bugs that are realistic enough to be used to improve language
model bugfixing capabilities. Here, we show that this is indeed the case through
fine-tuning language models on our benchmark dataset.

To measure bugfixing performance, we rely on two benchmarks: HumanEvalFix and
the Python program repair subset of MdEval. Canonical solutions in the HumanEval
benchmark are used in the construction of our benchmark dataset. As training on
these would inflate benchmark results, we remove the programs originating from
this benchmark from the fine-tuning dataset. Throughout this experiment, we
fine-tune Qwen2.5-Coder-0.5B-Instruct,
Qwen2.5-Coder-1.5B-Instruct~\citep{hui2024qwen25codertechnicalreport},
DeepSeek-Coder-1.3B-Instruct~\citep{guo2024deepseekcoderlargelanguagemodel} and
Yi-Coder-1.5B-Chat~\citep{yicoder}.

Each training sample starts with the corrupted program, followed by the
instruction to fix it and ends with the correct program as the response. We
only train the model on the output tokens, with the input tokens being masked.
Full parameter fine-tuning is used to adapt the models to the task of
bugfixing. The hyperparameters used for fine-tuning are consistent across all
models: the learning rate is set to \boldmath$1\cdot10^{-5}$, the effective
batch size is \textbf{8}, the warmup ratio is \textbf{0.05}, the \textbf{AdamW}
optimizer (adamw\_torch\_fused) is used, and each model is fine-tuned for one
epoch.

\begin{figure}[t!]
    \centering
    \vspace{0.5em}

    \begin{minipage}[b]{0.5\textwidth}
        \centering
        \begin{tikzpicture}
            \begin{axis}[
                ybar,
                bar width=10pt,
                ymin=0, ymax=60,
                ytick={0,10,20,30,40,50},
                height=6cm,
                width=\linewidth,
                enlarge y limits=0.1,
                symbolic x coords={
                    Qwen2.5-Coder-0.5B,
                    Qwen2.5-Coder-1.5B,
                    DeepSeek-Coder-1.3B,
                    Yi-Coder-1.5B,
                },
                enlarge y limits=0,
                enlarge x limits=0.2,
                xtick=data,
                xticklabel style={
                    rotate=15,
                    anchor=east,
                    font=\tiny,
                    xshift=2.5em,
                    yshift=-0.6em
                },
                xticklabels={
                    Qwen2.5-Coder-0.5B,
                    Qwen2.5-Coder-1.5B,
                    DeepSeek-Coder-1.3B,
                    Yi-Coder-1.5B
                },
                xlabel={Performance on HumanEvalFix},
                x label style={yshift=-0.5em},
                ymajorgrids=true,
                y grid style={dashed, gray!30},
                axis background/.style={fill=gray!10, opacity=0.5},
                nodes near coords,
                every node near coord/.append style={font=\scriptsize},
                point meta=explicit symbolic,
                legend style={
                    at={(1.2,1.1)},
                    anchor=south,
                    legend columns=-1,
                    font=\normalsize
                },
            ]
            \addplot[
                style={blue, fill=blue, opacity=0.6},
                every node near coord/.style={
                    font=\scriptsize,
                    rotate=90,
                    xshift=17pt,
                    yshift=0.00em,
                },
            ] 
            coordinates {
                (Qwen2.5-Coder-0.5B,12.20) [12.20\%]
                (Qwen2.5-Coder-1.5B,35.37) [35.37\%]
                (DeepSeek-Coder-1.3B,20.12) [20.12\%]
                (Yi-Coder-1.5B,26.83) [26.83\%]
            };
            \addplot[
                style={red, fill=red, opacity=0.6},
                every node near coord/.style={
                    font=\scriptsize,
                    rotate=90,
                    xshift=17pt,
                    yshift=-1.3em,
                },
            ] 
            coordinates {
                (Qwen2.5-Coder-0.5B,25.61) [25.61\%]
                (Qwen2.5-Coder-1.5B,39.63) [39.63\%]
                (DeepSeek-Coder-1.3B,26.83) [26.83\%]
                (Yi-Coder-1.5B,41.46) [41.46\%]
            };
            \legend{Before fine-tuning, After fine-tuning}
            \end{axis}
        \end{tikzpicture}
    \end{minipage}%
    \hfill
    \begin{minipage}[b]{0.5\textwidth}
        \centering
        \begin{tikzpicture}
            \begin{axis}[
                ybar,
                bar width=10pt,
                ymin=0, ymax=90,
                ytick={0,10,20,30,40,50,60,70,80},
                height=6cm,
                width=\linewidth,
                enlarge y limits=0.1,
                symbolic x coords={
                    Qwen2.5-Coder-0.5B,
                    Qwen2.5-Coder-1.5B,
                    DeepSeek-Coder-1.3B,
                    Yi-Coder-1.5B,
                },
                enlarge y limits=0,
                enlarge x limits=0.2,
                xtick=data,
                xticklabel style={
                    rotate=15,
                    anchor=east,
                    font=\tiny,
                    xshift=2.5em,
                    yshift=-0.6em
                },
                xticklabels={
                    Qwen2.5-Coder-0.5B,
                    Qwen2.5-Coder-1.5B,
                    DeepSeek-Coder-1.3B,
                    Yi-Coder-1.5B
                },
                xlabel={Performance on MdEval},
                x label style={yshift=-0.5em},
                ymajorgrids=true,
                y grid style={dashed, gray!30},
                axis background/.style={fill=gray!10, opacity=0.5},
                nodes near coords,
                every node near coord/.append style={font=\scriptsize},
                point meta=explicit symbolic,
            ]
            \addplot[
                style={blue, fill=blue, opacity=0.6},
                every node near coord/.style={
                    font=\scriptsize,
                    rotate=90,
                    xshift=15pt,
                    yshift=0.00em,
                },
            ] 
            coordinates {
                (Qwen2.5-Coder-0.5B,37.70) [37.70\%]
                (Qwen2.5-Coder-1.5B,49.18) [49.18\%]
                (DeepSeek-Coder-1.3B,37.70) [37.70\%]
                (Yi-Coder-1.5B,44.26) [44.26\%]
            };
            \addplot[
                style={red, fill=red, opacity=0.6},
                every node near coord/.style={
                    font=\scriptsize,
                    rotate=90,
                    xshift=15pt,
                    yshift=-1.3em,
                },
            ] 
            coordinates {
                (Qwen2.5-Coder-0.5B,47.54) [47.54\%]
                (Qwen2.5-Coder-1.5B,63.93) [63.93\%]
                (DeepSeek-Coder-1.3B,59.02) [59.02\%]
                (Yi-Coder-1.5B,62.30) [62.30\%]
            };
            \end{axis}
        \end{tikzpicture}
    \end{minipage}
    
    \caption{Bugfixing performance before and after fine-tuning on a subset of
    MegaBugFix. This experiment aims to train models on the corrupted programs
    and their corresponding correct variants in our dataset, with the goal of
    improving their bugfixing capabilities. The noticeable performance gains on
    HumanEvalFix and the Python bugfixing subset of MdEval indicate that
    MegaBugFix contains realistic, high-quality bugs.}
    \label{fig:finetune-results}
\end{figure}
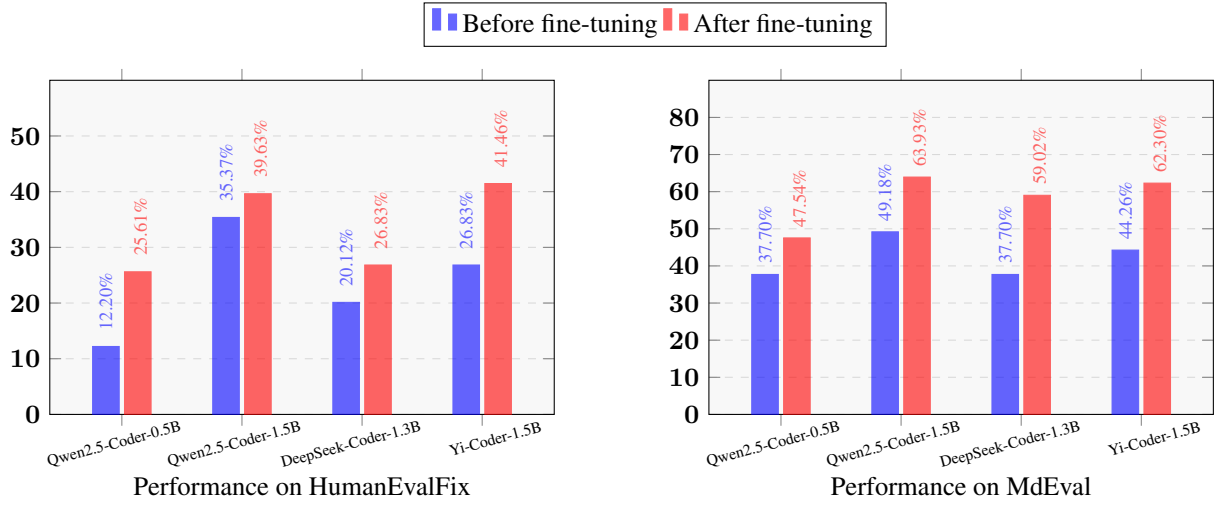

For evaluation, the same generation parameters are used as described in
\autoref{sec:comparison}. The outcomes of this experiment are presented in
\autoref{fig:finetune-results}, which visualizes bugfixing performance of both
the original models and their fine-tuned variants, as measured on two
benchmarks.

\section{Discussion}

We locally evaluated several open-weight LLMs on MegaBugFix, HumanEvalFix,
QuixBugs and MdEval (visualized in \autoref{fig:results}). We can observe from
these results that models consistently achieve lower performance on MegaBugFix
compared to other established benchmarks. This suggests that our benchmark
presents more difficult bugs to the models and uncovers failures that might
remain hidden when using prior benchmarks. The gap between benchmark results is
particularly noticeable for example in case of the CodeGemma-7B-Instruct and
Qwen2.5-Coder-1.5B-Instruct models, which perform significantly better on
established benchmarks than on MegaBugFix. The higher difficulty of the
benchmark likely comes from the LLM used for bug injection, which was trained on
real-world, more complex bugs. Furthermore, it can also be the result of our
dataset being larger and more diverse.

To assess the quality of the bugs in the benchmark, we fine-tuned LLMs on its
dataset to see if they can improve in bugfixing performance
(\autoref{fig:finetune-results}). We can see that all models improve
considerably after fine-tuning them on our proposed dataset. Overall, the
results indicate that the corrupted programs in our benchmark do reflect
realistic bugs, as learning to fix bugs from them improves performance.

\section{Threats to Validity}

A potential threat to validity of our study is related to the experiment
designed to improve bugfixing performance through fine-tuning on our benchmark
dataset. In this experiment, we relied exclusively on language models below 1.5B
parameters. Consequently, our findings may not generalize to larger models.

Furthermore, in our code corruption pipeline, we used seemingly arbitrary
filtering thresholds and did not experiment with alternative values. Even though
these intuitively chosen thresholds proved sufficiently robust for our use case,
further experiments should be conducted in this regard.

Finally, we evaluated models only up to 32B parameters and excluded
closed-weight ones from the evaluation pipeline. These factors might constrain
the generalizability of the baseline performance results on our benchmark.

\section{Conclusion}

In this paper, we presented MegaBugFix, a large-scale benchmark to evaluate
bugfixing capabilities of tools such as Large Language Models. By fine-tuning an
open-weight Large Language Model for automatic bug injection via diff
generation, we obtained 12,629 buggy variants of correctly implemented programs,
forming the core of the proposed benchmark. The original correct programs
originate from existing benchmarks and datasets, and are obtained with their
corresponding test cases. We developed and published a unified framework to
measure bugfixing capabilities on the corrupted programs by running the test
cases. We hope that the ability to utilize this framework will aid the research
community in evaluating and comparing various bugfixing approaches in the
future.

For future work, we plan to improve MegaBugFix in two aspects. On one hand, we
aim to support more programming languages beyond Python. On the other hand, we
will expand the coverage from bug fixing alone to a broader range of
bugfix-related tasks, such as bug localization, identification, and analysis,
to evaluate more aspects of the program repair workflow.

\section*{Acknowledgements}

Supported by the EKÖP-25 University Excellence Scholarship Program of the
Ministry for Culture and Innovation from the source of the National Research,
Development and Innovation Fund.

\newpage

\bibliographystyle{acm}
\bibliography{references}

\end{document}